# Path-Integrated Polarization Rotation Signatures in Subsea Networks: Cable Geometry Effects in 2023 Turkey Earthquakes

Mohammad M. Hosseini(1), Miquel Masanas(1), Giuseppe Parisi(1), Wesley Braga Melo(3), Alberto Marullo(2), Danilo Decaroli(2), Sergei K. Turitsyn(3), Antonio Napoli(1)

(1) Nokia, Optical Networks, Munich, Germany, mohammad.hosseini@nokia.com, (2) Sparkle,Roma, Italy, (3) Aston University, Birmingham, UK

***Abstract*** *We analyze polarization data streamed from a commercial coherent transponder, used in MedNautilus link, during 2023 Turkey earthquakes. We show how earthquake features and cable geometry fundamentally govern the optical polarization response, taking the first steps for detailed interpretation in designing fiber-based seismic sensing systems. ©2026 The Author(s)*

## Introduction

The utilization of submarine fiber-optic cables for seismic sensing is rapidly evolving from a niche experimental concept into a transformative tool for global geophysics. In coherent optical transponders, monitoring the State-of-Polarization (SOP) has emerged as a promising sensing modality, as the fiber birefringence evolves in response to environmentally induced strains [3] and shows its time evolution in the Jones matrices or, alternatively, Stokes parameters. While these links can capture seismic perturbations over thousands of kilometers, in non-reflective measurements the sensing paradigm is intrinsically shifted, because the measurement provides a path-integrated response that reflects the cumulative effect of perturbations along the entire one-dimensional optical fiber cable. Consequently, the measured SOP is not a spatially resolved observation of the cable strain, but a compressed projection of heterogeneous interactions, including source radiation patterns, seismic wave polarization, propagation geometry, and cable–seafloor coupling. This path-integrated character presents a significant interpretive bottleneck: it is exceptionally difficult to disentangle the true nature of the seismic source from the geometric biases of the sensing cable. Understanding these effects is essential for deploying coherent transponders as reliable sensing elements in operational submarine networks. Motivated by this interpretive challenge, we consider the 2023 Turkey earthquake doublet which provides an unprecedented natural laboratory to investigate this challenge. Comprising a prolonged, multi-segment subshear Mw 7.8 event (80 seconds) and a highly asymmetric, supershear Mw 7.5 event (40 seconds) [4, 5], the doublet presents two different events whose waves interacted with the exact same Mediterranean fiber route enabling us to investigate optical fiber biases enabling a better understanding of the strengths and limitations of this technique.

We developed a framework to simulate path-integrated strain caused by pair of earthquakes and cross-analyzed them with real-world, compensated rotation-vector measurements extracted from a coherent transponder on the subsea cable MedNautilus from Catania to Haifa (operated by Sparkle). Instead of treating the fiber as a point sensor, we unroll the cable route and model incoming waves as localized strain components. By combining a vectorial strain-projection model with 2D cylindrical wave propagation, and source-specific radiation patterns, we track how local ground motion transforms into an aggregated polarization perturbation signal. Our goal is to joint analyze the earthquake size, wave polarization, and cable geometry and length, establishing guidelines for using telecom infrastructure in seismic monitoring. We show cable geometry creates deterministic sensitivity patterns, and can have blind spots and higher sensitivity to Rayleigh waves.

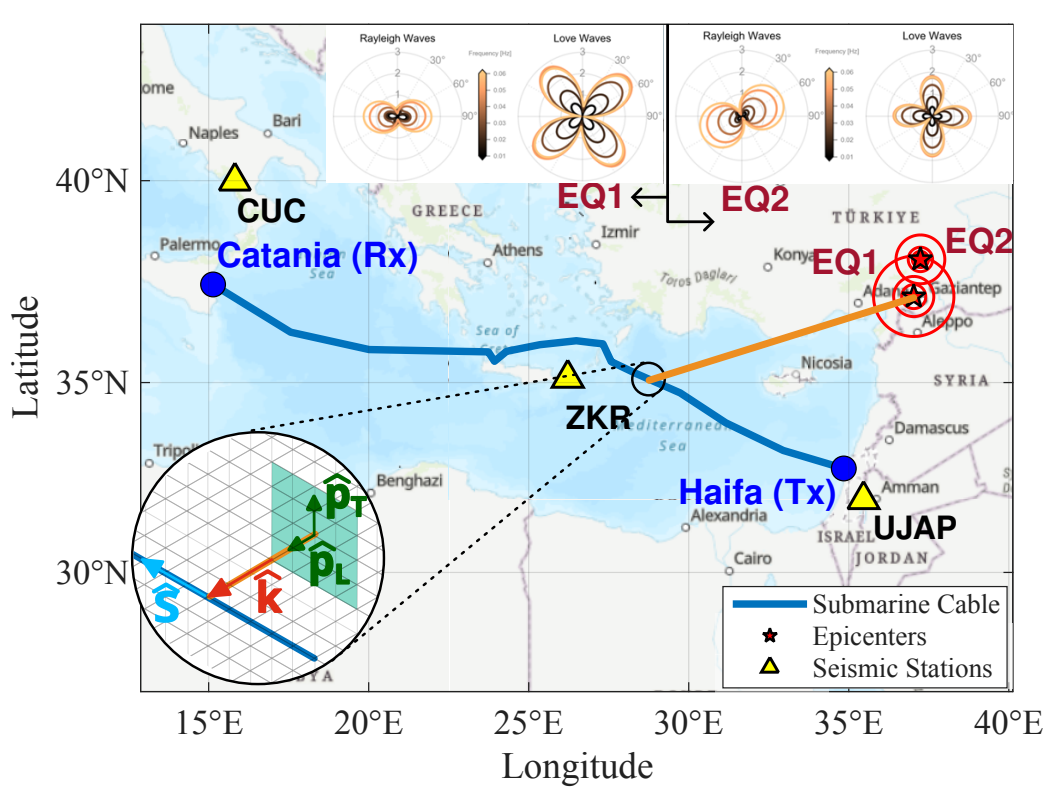

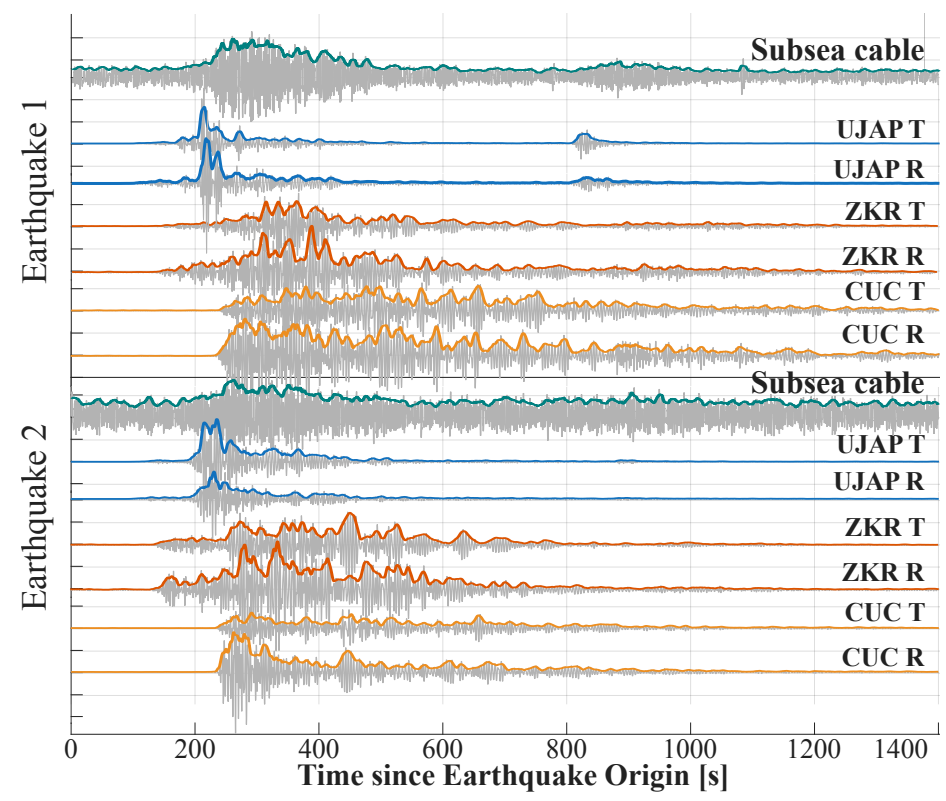

**Fig. 1:** (a) Study framework including the cable route (around 2300 km flat route), 3 seismic stations [1], earthquakes epicenter, and their surface wave radiation patterns [2] (b) the filtered waveforms (0.2 Hz-0.5 Hz) for optical polarization and seismogram.

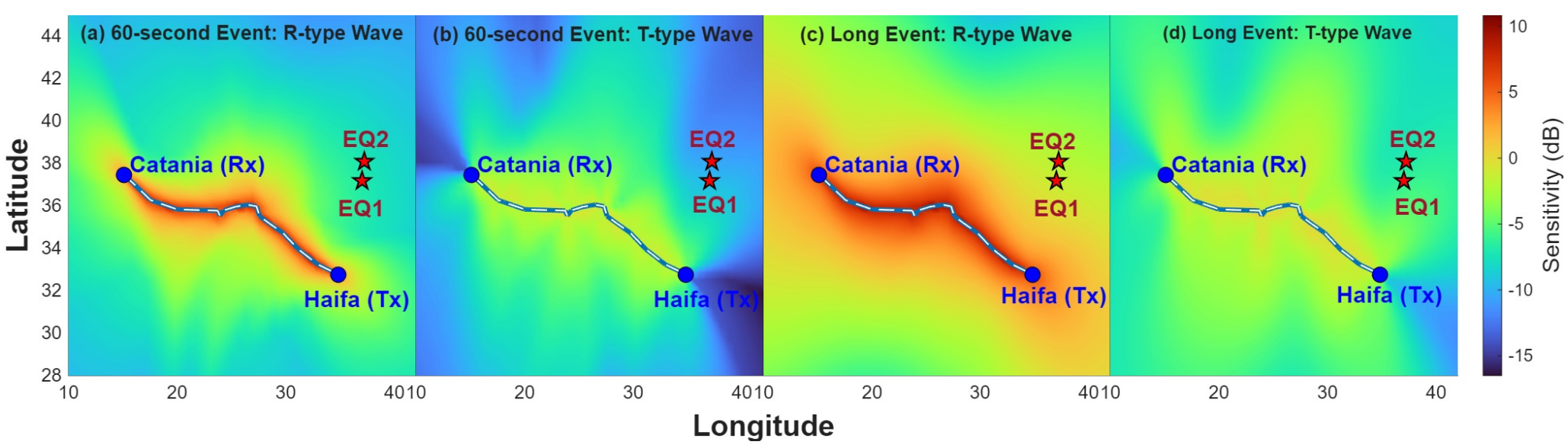

**Fig. 2:** Maximum geometrical sensitivity $|\Gamma|^2$ of the Catania–Haifa cable based on Eq. 1 according to the epicenter location, assuming 3 km/s wave speed highlighting geometry-induced sensitivity variations.

## Theoretical Framework

Unlike conventional seismometers that record omnidirectional particle velocity at a discrete point, a submarine fiber-optic cable acts as a distributed, one-dimensional (1D) strainmeter. It measures dynamic axial strain ($\epsilon_{ax}$), physically projecting the local ground deformation onto the 1D axis of the fiber. Consequently, its directional sensitivity acts similarly to a dipole antenna, exhibiting distinct blind spots dictated by the wave's incidence angle and its polarization [6, 7]. While a traditional seismogram captures localized ground motion, a coherent transponder provides the SOP—an aggregated signal integrating perturbations spread across the total length of the link.

To quantify the local axial strain, we define three fundamental unit vectors: the local fiber segment orientation ($\hat{\mathbf{s}}$), the seismic wave propagation direction ($\hat{\mathbf{k}}$), and the earthquake wave polarization direction ($\hat{\mathbf{p}}$). The strain sensed by the fiber depends on its alignment with both the wave propagation direction and the particle motion, and can be expressed as the product of two scalar dot products as $\epsilon_{ax} \propto (\hat{\mathbf{s}} \cdot \hat{\mathbf{k}})(\hat{\mathbf{s}} \cdot \hat{\mathbf{p}})$, assuming a monochromatic plane wave (i.e., single-frequency, ignoring spectral and wavefield complexity) [8]. This vectorial formulation natively resolves the directional sensitivity for both radial waves (R-type, where $\hat{\mathbf{p}} \parallel \hat{\mathbf{k}}$) and transverse waves (T-type, where $\hat{\mathbf{p}} \perp \hat{\mathbf{k}}$) without requiring complex tensor rotations.

The geometrical sensitivity factor can be obtained by spatially integrating squared strain amplitudes assuming unit wave amplitude at source and unit strain-to-birefringence factor. $|\Gamma_{R,T}|^2$ for R and T waves is scaled by the square of the local vector projection and by the source's amplitude radiation pattern $\mathcal{R}(\theta)$ and path loss due to distance $A(r)$ (which is $1/\sqrt{r}$ for a 2D wave where $r$ is the great circle distance from the epicenter to the incident point with the cable). Note that squared magnitude fluctuation of compensated polarization rotation vector $|\Delta\phi|^2$ is proportional to $|\Gamma_{R,T}|^2$. The sensitivity for R- and T-type waves is proportional to the sum of squared projected local strain over the continuous cable path $s$ at a given time:

$$|\Gamma_{R,T}|^2 = \int A(r)^2\, \mathcal{R}_{R,T}(\theta)^2 \left[(\hat{\mathbf{s}} \cdot \hat{\mathbf{k}})(\hat{\mathbf{s}} \cdot \hat{\mathbf{p}}_{R,T})\right]^2 ds \qquad (1)$$

where $\theta$ is the departure azimuth. While Eq. 1 utilizes a continuous spatial integral ($ds$), the accumulation of the measured Jones Rotation vector is physically discretized by the fiber's birefringence correlation length ($L_c$). Because transoceanic distances vastly exceed this length ($L_{cable} \gg L_c$), the cable acts as a sequence of independent, randomly oriented waveplates. Therefore, the polarization rotation axis direction undergos a random walk along the fiber, and the environmental perturbations they experience are de-correlated through this mechanism. This leads to incoherent signal accumulation, and therefore causes that the changes in polarization rotation vary their variance with integrated squared local strain (projected to the fiber), rather than the coherent addition of its amplitude. Phase metrology [9] [10], on the other hand, has coherent integration of the perturbations along the cable. While incoherence accumulation prevents straightforward localization, unlike in distributed acoustic sensing, neither SOP or phase metrology can be spatially differentiated in forward-only measurements and the incoherent accumulation becomes a source of resilience against positive and negative strains canceling out.

For improved precision, cross-correlation can be applied to seismogram data after rotating the recorded North, East, and vertical components into a physically meaningful coordinate system aligned with the direction of the wave propagation. The waveforms (or their envelopes) from two different sensors are then cross-correlated in time to determine the lag at which the signals best align. The time shift corresponding to the maximum correlation indicates the time that the wave travels the differential distances. Applying this procedure within specific frequency bands allows the extraction of group velocities even in the presence of significant background noise.

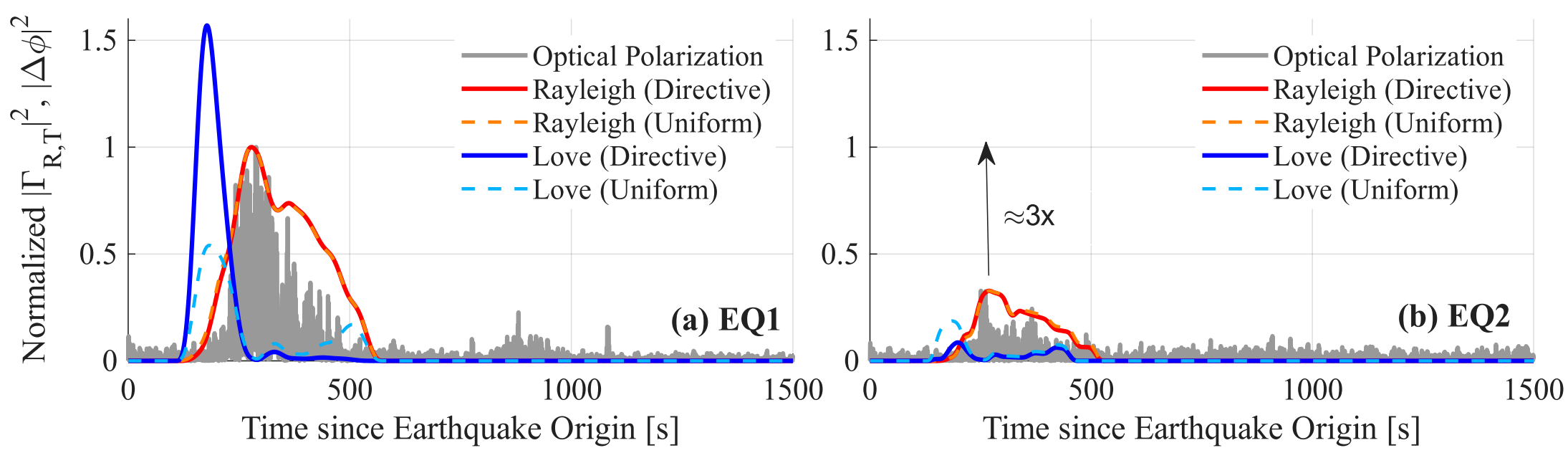

**Fig. 3:** Comparison of polarization response with the surface waves models with realistic and uniform radiation patterns.

## Results and Discussion

Data from three Mediterranean seismic stations were paired with $20$ Hz Jones matrix data from a coherent transponder at the Catania terminal [11]. Following [12], raw Jones data were converted to polarization rotation vectors and drift-compensated using a $2$ s moving average. Seismic signals, sampled at $20$ Hz or $40$ Hz across three orthogonal channels, were projected onto the propagation path to resolve R-type and T-type wave components. Also all signals have been filtered between 0.2 to 0.5 Hz where the background noise is lower [13] and to have a common baseline (we found out the polarization response to the second earthquake is detectable in this band). All waveforms and their envelopes are shown in Fig. 1 (for subsea cable magnitude of the rotation vector which is signed based a given reference orientation). The closest station to the epicenters is UJAP (≈700 km) while CUC (≈1800 km) is the furthest. In terms of arrival of time, the optical response is most similar to the ZKR station.

We model the source with a squared-sine envelope, $A(t) = \sin^2\left(\frac{\pi t}{T}\right)$ for $t \in [0, T]$, yielding a smooth waveform that vanishes at both endpoints. To compute the $|\Gamma|^2$ over a source duration $T$, the wave source is modeled as radially propagating forming a disk-shaped wavefront. This geometry is characterized by an outer radius $R_2$ and an inner radius $R_1$, related through the propagation velocity $v$ by $R_2 - R_1 = vT$. Fig. 2 presents the sensitivity for R and T waves as the epicenter varies over $28° - 45°$ latitude and $10° - 42°$ longitude under a uniform radiation pattern for a short-duration source (active segment $R_2 - R_1 < L_{cable}$) and long-duration sources (entire cable interacting $R_2 - R_1 > L_{cable}$ at once). The cable is divided into 10000 segments for the simulation. The R waves exhibit higher sensitivity, while T waves show blind spots, as the nearly linear fiber geometry allows $\hat{k}$ to align with the fiber over extended segments. Longer events, which interact with the entire cable length at once, make the sensitivity more uniform.

We decomposed the seismogram data to T-waves and R-waves by projecting on $\hat{k}$ and the perpendicular direction to it. Unlike Rayleigh waves, which involve coupled vertical and longitudinal motion, Love waves are trapped surface waves that exhibit purely transverse motion [14]. Assuming flat 2D layout of fiber, the Rayleigh is mainly R-type and the Love waves are T-waves. Using seismic stations signals cross-correlation, we calculated the speed of waves as $v_{R1} = 3.6$, $v_{T1} = 3.6$, $v_{R2} = 3.8$, and $v_{L2} = 4.3$ km/s. Next, we simulated the $|\Gamma|^2$ obtained using velocity values, source modeling, radiation patterns and Eq. 1. We used Rayleigh and Love wave radiation patterns from Incorporated Research Institutions for Seismology (IRIS) can be accessed through IRIS [2]. Fig. 3 shows normalized $|\Delta\phi|^2$ for both earthquakes, and $|\Gamma|^2$ both wave types considering uniform and non-uniform radiation. Note that the second earthquake is normalized to the maximum of the first earthquake's $|\Delta\phi|^2$, while the Love waves are normalized to the peak Rayleigh-wave sensitivity $|\Gamma_R|^2$ all scaled by $|\Delta\phi|^2$. The relative energy for these two earthquakes is $\frac{E_1}{E_2} = 10^{\frac{3}{2}(7.8-7.5)} \approx 2.8$ which is comparable to the relative peak of $|\Delta\phi|^2$ for two events (the distances to cable are similar). As can be seen, Rayleigh waves match well with the optical signal obtained using a coherent transponder for these two events. Our analysis reveals that while the onset of Love and Rayleigh waves is similar, the main energy peak for Love waves is concentrated in the first segment of the optical fiber, from Haifa to around the midpoint, whereas the Rayleigh waves exhibit a later peak highlighting the geometric orientation sensitivity of the fiber-optic cable, which is influenced by the way the waves interact with the ground. However, the strong match with Rayleigh waves suggests that the first segment of the cable might have poor seabed coupling, or that the real cable's varying topography makes it less sensitive to Love waves compared to Rayleigh waves.

## Conclusion

Submarine fiber-optic cables can serve as massive strain integrators, providing a cost-effective means of global seismic monitoring. We showed the relationship between integrated polarization, cable geometry, wave types, and seismic features; simulations indicated that the measurements match the relative sizes of earthquake magnitudes and Rayleigh wave interaction.

## Acknowledgment

This work has been performed in the framework of the ECSTATIC project, which received funding from the European Union's Horizon Europe Framework Programme under grant agreement No 101189595. Miquel Masanas research leading to these results has received funding from the European Union's Horizon Europe research and innovation programme under Grant Agreement No. 101095055 – SUBMERSE. We express our best gratitude to Dr. Susana Custodio, Dr. Luis Matias (U Lisboa) and Dr. Susana Oliveira (INESCTEC) for their contribution in geophysical analysis and understanding. Waveform data were obtained via IRIS Web Services using the irisFetch MATLAB client. We acknowledge EarthScope Data Services, GEOFON, and INGV, as well as the contributing FDSN networks, for providing open access to seismic data.